\documentstyle[prl,aps,multicol,epsfig,epsf]{revtex}
\large
\begin{document}
\draft
\title{Partition noise and statistics in the fractional quantum
Hall effect} 
\author{I. Safi$^1$, P. Devillard$^{1,2}$ and T. Martin$^{1,3}$}
\address{ $^1$ Centre de Physique Th\'eorique, 
Case 907 Luminy, 13288 Marseille Cedex 9, France}
\address{$^2$ Universit\'e de Provence, 13331 Marseille Cedex
03, France} 
\address{$^3$ Universit\'e de la M\'editerran\'ee, 
13288 Marseille Cedex 9, France} 
\maketitle
\begin{abstract}
A microscopic theory of current partition in fractional
quantum Hall liquids, described by chiral Luttinger liquids, 
is developed to compute the noise correlations, using the Keldysh
technique. In this Hanbury-Brown and Twiss geometry,
at Laughlin filling factors $\nu=1/m$, the real time
noise correlator exhibits oscillations which persist over larger
time scales than that of an uncorrelated Hall fluid. The zero
frequency noise correlations are negative at filling factor
$1/3$ as for bare electrons (anti-bunching), but are strongly
reduced in amplitude. These correlations  become positive
(bunching) for $\nu\leq 1/5$, suggesting a tendency towards
bosonic behavior.   \end{abstract}  \begin{multicols}{2}  
\narrowtext

\pacs{PACS 72.70+m,71.10.Pm,73.40.Hm}

Transport experiments in the fractional quantum Hall effect
(FQHE) \cite{laughlin} have provided a direct measurement of the
fractional charge of the quasi-particles
\cite{saminad,picciotto} associated with these  correlated
electron fluids. These results  constitute a preliminary test of
the Luttinger liquid  models
\cite{kane_fisher_noise,chamon_noise} based on chiral edge
Lagrangians \cite{wen} which describe the low-lying
edge excitations. However, the discussion has centered on 
the charge of the quasiparticles, rather than the statistics. On
the other hand, noise correlation experiments
\cite{henny,oliver} in branched mesoscopic  devices, i.e.
fermion analogs of the Hanbury--Brown and Twiss experiments for
photons \cite{hbt}, have detected the negative  noise
correlations predicted by theory   \cite{martin_landauer}.
Statistical features in transport are quite explicit
in such experiments. So far in the FQHE, the measurement of 
the noise reduction \cite{saminad} -- smaller than that of
fermions -- constitutes the only hint that the statistic
is not fermionic.

\begin{figure}
\epsfxsize 8.5 cm
\centerline{\epsffile{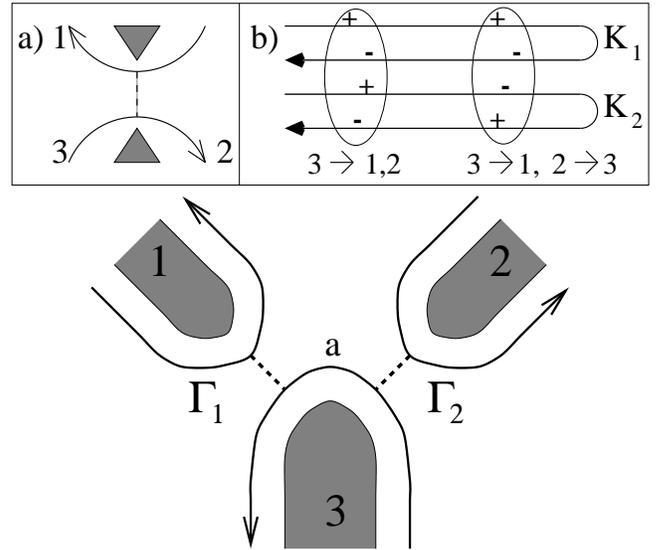}}
\medskip
\caption{\label{fig1}
Hanbury-Brown and Twiss geometry: $3$ metallic gates (grey)
define $3$ edge states, with $2$ tunneling paths separated by 
$a$ and with tunneling amplitudes $\Gamma_1$ and $\Gamma_2$.
Insets: a) Correlation experiment in a $2$-edge quantum Hall bar
with a  constriction. b) Charge configurations on the two Keldysh
contours $K_1$ and $K_2$ which contribute to the $\omega=0$
noise  correlations at finite (left and right) and at zero (left
only) temperature.} \end{figure}

Here, it is suggested that the statistics of the
underlying excitations of the FQHE can be monitored via a
Hanbury--Brown  experiment
where quasiparticles emitted from one edge and tunneling 
through the correlated Hall fluid are
collected into two receiving edges (see Fig. \ref{fig1}).
This constitutes a mesoscopic analogue of a collision 
process which involves many quasi-particles, and therefore 
provides a direct probe of their underlying statistics.
The Luttinger edge state theory
\cite{wen} is used to compute the current and noise 
with the Keldysh technique.
The analytic results for the noise in this partition
experiment show that: a) upon increasing the magnetic field
from the integer quantum Hall effect (IQHE) to filling factor
$1/3$, the (negative) correlations are strongly reduced
in amplitude; 
b) these correlations change sign and are positive at $\nu\leq
1/5$.  This work attempts  to go further than a recent proposal 
where statistics and scattering properties were dissociated 
\cite{isakov}, which correlations are fermionic 
\cite{torres_martin}. 

The suggested geometry is depicted in Fig.
\ref{fig1}: it requires {\it three} edges (two of which are
assumed to be decoupled), in contrast to previous
noise correlation measurements \cite{saminad,henny} in the IQHE
and in the FQHE where a single constriction controlled the
transmission between two edge states. There, negative noise 
correlations between the receiving ends of 
two edge states (inset Fig \ref{fig1}a) 
are the consequence of a noiseless 
injecting channel together with current conservation, 
for arbitrary $\nu$. 
On the other hand, removing excitations 
from one edge state and redistributing them to 
two other edges (Fig.\ref{fig1}) is clearly  
relevant for uncovering the 
bunching/anti-bunching properties of quasi-particles.
This setup can be considered as a detector of 
partition noise between  edge $1$ and $3$, but
in the presence of a ``noisy'' injecting current (due to 
backscattering between $2$ and $3$). 

The edge modes 
running along each gate, 
characterized by chiral bosonic fields $\phi_l$ ($l=1,2,3$)
are described by a  
Hamiltonian 
$H_{0}=(v_F\hbar/4\pi)\sum_{l=1,2,3}\int ds
(\partial_{s}\phi_l)^2$ with $s$ the curvilinear abscissa, 
and with a
current $v_F\sqrt{\nu}\partial_{s}\phi_l/2\pi$, which is
conserved in the absence of scattering ($v_F$ is the Fermi
velocity).  
$\phi_l$ satisfy the commutation relation
$[\phi_l(s),\phi_{l'}(s')]=i\pi \delta_{ll'}sgn(s-s')$
\cite{geller_loss}.   
The quasi-particle operators are
expressed as $\psi_l^\dagger(s)=(2\pi\alpha)^{-1/2}F_le^{ik_F s}
e^{i\sqrt{\nu}\phi_l(s)~}$,
where $\alpha$ is a cutoff.
Both the above commutation relation and the (unitary) Klein
factors $F_l$ guarantee fractional exchange statistics provided
that:   \begin{equation}
F_lF_{l'}=e^{-i\pi p_{ll'}\nu}F_{l'}F_l
\label{klein}
\end{equation}
with $3$ possible statistical phases $p_{ll'}=-p_{l'l}=\pm
(1-\delta_{ll'})$. In particular, this insures that the fields
$\psi_l$  anti-commute for  $\nu=1$ and commute for
$\nu\rightarrow 0$.

Tunneling of quasi-particles occurs at two locations 
$s=\pm a/2$ on edge $3$, and at $s=0$ on edges $1$
and $2$. 
A non--equilibrium situation is achieved by imposing a 
bias $\hbar\omega_l/e^*=-\partial\chi_l/\partial t$ ($l=1,2$)
between $3$ and $l$, where $\chi_l$ denotes the gauge parameter
which appears in the tunneling Hamiltonian
$H_B=H_{B1}+H_{B2}$, with: 
\begin{equation}
H_{Bl}=\Gamma_le^{-ie^*\chi_l/\hbar
c}\psi_l^\dagger(0)\psi_3 \left((-1)^la/2\right)+H.c.
~.\label{tunneling Hamiltonian} \end{equation} 
$H_{B1}$ and $H_{B2}$ are required to commute \cite{nayak}, 
which imposes the constraint $p_{12}+p_{23}+p_{31}=1$ on the
statistical phases of Eq. (\ref{klein}).

Non--equilibrium averages are 
extracted from the Keldysh partition
function. The perturbation 
theory is analogous to the Coulomb
gas models which have been proposed to study transport
in Luttinger models \cite{chamon_noise}. Here two contours
$K_l$ ($l=1,2$) contain $m_l$ (even) charges ($\pm$)
which account for the quasi-particle transfer to/from edge 
$3$ to $l$ at time $t_{lk}$ attached to the upper/lower branch
of $K_l$.  Expanding the
exponential, the partition function reads:
\begin{eqnarray}\label{partition} 
&&Z_K=\sum_{m_1,m_2=0}^\infty \left({-i\over
\hbar}\right)^{m_1+m_2}\sum_{\mathcal{C}} \int\!
{dt_{11}...dt_{1m_1}dt_{21}...dt_{2m_2} \over m_1! m_2!}
\nonumber\\ && \times\langle T_K
H_{B1}(t_{11})...H_{B1}(t_{1m_1})
H_{B2}(t_{21})...H_{B2}(t_{2m_2})\rangle_{\mathcal{C}} ~.
\end{eqnarray} where the subscript $\mathcal{C}$ identifies
charge  configurations. Relevant charge
configurations which contribute to lowest order to the noise
correlations are depicted  in Fig. \ref{fig1}b): charge
neutrality is imposed on each contour. 
The terms in the nonequilibrium average
Eq. (\ref{partition}) can in
general be decoupled into four contributions: one Keldysh
ordered product of bosonic fields for each edge, which dynamics
are specified by the Green's function of the chiral boson fields
\cite{chamon_freed} $G_{\eta\eta'}(s,t)$
(with contour branch labels $\eta,\eta'=\pm$), and a fourth 
contribution which arises from the Klein factors,  which have no
dynamics of their own, yet which  are essential in order to
specify the tunneling operators.     

\begin{figure}[htb] \begin{center}
\mbox{\epsfig{file=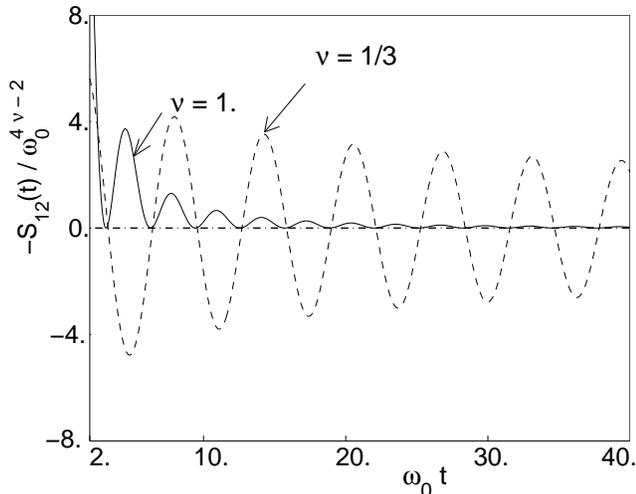,width=8.5cm}}
\narrowtext
\caption[to]{
Contribution to 
the real time correlator $S_{12}^{-+}(t)$, for
bias $\omega_0=e^*V/\hbar$, normalized to  
$|\tau_0\Gamma_1 \Gamma_2|^2|\omega_0|^{4\nu-2}$ for a
filling factor $\nu=1/3$ (dashed line). 
$2\langle I_1\rangle\langle I_2\rangle$ has
been subtracted. The
exact {\it excess} noise at $\nu=1$:
$S_{12}^{ex}(t)\propto \sin^2(\omega_0t/2)/t^2$ 
(full line), keeping all $\eta$ configurations, is  plotted
for comparison.}  \label{fig2} \end{center}
\end{figure}

The quasi-particle current operator between leads $3$ and
$l=1,2$ is  $I_l= -c\,\partial H_B/\partial \chi_l$. The
symmetrized real time current--current correlator between edges
$1$ and $2$ which is used to compute the noise contains two time
arguments, which are assigned to different branches of the
Keldysh contour (thus the notation $\chi_l^\eta(t)$ below).
Performing functional  derivatives on $Z_K$,     
\begin{equation} S_{12}(t-t')=-(\hbar c)^2\sum_{\eta=\pm}
{\delta^2 Z_K\over \delta \chi_1^\eta(t)
\delta\chi_2^{-\eta}(t')}\biggl|_{\omega_{1,2}=\omega_0}~,
\label{Keldysh time correlator}
\end{equation} 
assuming equal biases on $1$ and $2$.
The leading term in Eq. (\ref{Keldysh time correlator})
is of fourth order in the tunneling 
amplitudes,  corresponding to
$m_1=m_2=2$ in Eq. (\ref{partition}), in contrast to the leading
contribution to the individual currents and noises (second order
``dipole'' contributions).  Exploiting the symmetry property of
the Green's function
$G_{-\eta,-\eta'}(s,t)=\left[G_{\eta,\eta'}(s,t)\right]^*$: 
\begin{eqnarray}
\label{real time Sbis}
S_{12}(t)&=&4\frac{|e^*\tau_0\Gamma_1\Gamma_2|^2}
{(h\alpha)^4}Re
\int_{-\infty}^{\infty}\!\!dt_1\int_{-\infty}^{\infty}\!\!dt_2
\sum_{\epsilon,\eta_1,\eta_2=\pm}\epsilon\eta_1\eta_2
\\\nonumber &&
\times \cos\left(\omega_0(t_1+\epsilon t_2)\right)
e^{2\nu\left[G_{+,\eta_1}(0,t_1)+G_{-,\eta_2}(0,t_2)\right]}
\\\nonumber && \times
\frac{e^{\nu\epsilon\left[\widetilde{G}_{+\eta_2}(-a,t+t_2)
+\widetilde{G}_{\eta_1,-}(-a,t-t_1)\right]}}{e^{\nu\epsilon
\left[\widetilde{G}_{+-}(-a,t)
+\widetilde{G}_{\eta_1\eta_2}(-a,t+t_2-t_1)\right]}}~,
\end{eqnarray} 
where $\epsilon$ represents
the product of the two charge transfer processes: 
$\epsilon=-/+$ when the quasiparticles tunnel in the 
same/opposite
direction (left/right hand side of Fig. \ref{fig1}b).   
In Eq. \ref{real time Sbis}, the Green's 
function for edge $3$, which mediates the  
coupling between $K_1$ and $K_2$, has been 
translated due to the Klein factors:
$\widetilde{G}_{\eta\eta'}(-a,t)=G_{\eta\eta'}
(-a,t)+i\pi/4\left[(\eta+\eta')sgn(t)-\eta+\eta'\right]$. 
The integrand in the double integral in Eq. (\ref{real time
Sbis}) for $\nu<1$ decays slowly with both time arguments.  At
large times, the last factor in the integrand is equal to $1$,
thus corresponding to the product of the current  averages
$2\langle I_1\rangle\langle I_2\rangle$. Absolute convergence is
obtained for $\nu>1/2$ from the power law decay in time.  For
$\nu<1/2$ convergence  is due to the oscillatory terms.   
Zero temperature, $a=0$, and a symmetric bias
$\omega_{1,2}=\omega_0$ are chosen in order to enhance
statistical signatures.  Experimentally this implies that
the two tunneling paths lie within a few Fermi wave-lengths
from each other. The overlap between 
quasiparticles in edge $3$ 
is then more prominent. 
Here, only  
$\eta_2=-\eta_1=1$ is retained because first, it provides the 
large time behavior and second, it corresponds to the
contribution of the zero frequency noise correlations (to be
computed later on). Using  the explicit expressions of the
Green's function at equal abscissa, 
$G_{\eta\eta'}(t)=-\ln[1+it(\eta\theta(t)-\eta'\theta(-t))/\tau_0]$ 
and rescaling the
times by the short time cutoff $\tau_0\sim \alpha/v_F$, this
contribution reads:     
\begin{eqnarray}\label{real time Sbisbis}
&&S_{12}^{-+}(t)=-4\frac{|e^*\tau_0\Gamma_1\Gamma_2|^2}
{(h\alpha)^4}Re
\int_{-\infty}^{\infty}\!\!dt_1\int_{-\infty}^{\infty}\!\!dt_2
\sum_{\epsilon=\pm}\epsilon  \nonumber\\
&& \times \cos\left[\omega_0(t_1+\epsilon t_2)\right]
\left(1-it_1\right)^{-2\nu}
\left(1+it_2\right)^{-2\nu}
(1-it)^{\epsilon\nu}\nonumber\\
&& \times
\left[1+i(t+t_2-t_1)\right]^{\epsilon\nu}
\left(1+i|t+t_2|\right)^{-\epsilon\nu}
\left(1-i|t_1-t|\right)^{-\epsilon\nu}\nonumber\\
&& \times
\exp\left(i\epsilon\frac{\pi}2\nu\left[sgn(t+t_2)-sgn(t-t_1)\right]\right).
\end{eqnarray}
A leading contribution to $S_{12}^{-+}(t)$ is
plotted in  Fig. \ref{fig2}, as well as the excess noise at
$\nu=1$ for comparison. The latter oscillates with a frequency
$\omega_0$, and decays as  $t^{-2}$. $S_{12}^{-+}(t)$   scales as
$|\omega_0|^{4\nu-2}f(\omega_0 t)$, with $f(x)$ an oscillatory
function which decays at least as  $x^{-2\nu}$, thus a slower
decay than that of electrons. At large times, the frequency
of the oscillations stabilizes as $\omega_0=e^*V/\hbar$.  

\noindent
The result in Eq. (\ref{real time Sbis}) is now integrated over
$t$ after subtracting the average current products:
$\tilde{S}_{12}(\omega)\equiv \int dt\, e^{-i\omega
t}[S_{12}(t)-2\langle I_1\rangle \langle I_2\rangle]$.
The sign and magnitude of the $\omega=0$ correlations tell us 
the tendency  for the quasiparticle to exhibit bunching or
antibunching. Turning now to the charge configurations of Fig. 
\ref{fig1}b, at zero temperature only $\epsilon=-1$ in 
$\tilde{S}_{12}(0)$ contributes, which gives the 
information that an ``exclusion principle'' prohibits the
excitations to be transfered from the collectors to the emitter. 
The zero frequency noise correlations have the general form:
\begin{equation}
\tilde{S}_{12}(0)=  (e^{*2}|\omega_0|/ \pi) T_1^rT_2^r R(\nu)
\label{general correlation}
\end{equation}
where the renormalized transmission probabilities are
$T_l^r=(\tau_0|\omega_0|)^{2\nu-2}\left[\tau_0 \Gamma_l/\hbar
\alpha\right]^2/\Gamma(2\nu)$, and the dimensionless function
$R(\nu)$ characterizes the statistical correlations. 
At $\nu=1$, it is shown explicitly that $R(1)=-1$
using contour integration, so
that $\tilde{S}_{12}$ coincides  exactly with the
scattering theory result \cite{martin_landauer}. This 
issue represents a crucial test of the implementation of the
Klein factors.  Moreover, for arbitrary $\nu$, $R(\nu)$ could
in principle be directly measured in an experiment.
Indeed one can rescale the noise correlation
$\tilde{S}_{12}$ by the individual shot noises 
$\tilde{S}_l\simeq 2e^*\langle I_l\rangle$ or equivalently
(at this order) by the individual currents:
$R(\nu)=|\omega_0|\tilde{S}_{12}/[4\pi\langle
I_1\rangle\langle I_2\rangle]$.

A central
result of this letter is the analytical expression for the
function $R(\nu)$ in Eq. (\ref{general correlation}). It is
obtained by performing a change of variables of the $3$
time integrals in $\tilde{S}_{12}(0)$, neglecting the
short-time cutoff in the diagonal elements of the Keldysh
Green's function in Eq. (\ref{real time Sbis}). This procedure
is consistent with the limit of small biases which 
is assumed here $|\omega_0|\tau_0\ll 1$, but strictly speaking
it is valid in the range $1/2<\nu\leq 1$. One then obtains 
the asymptotic series:    
\begin{eqnarray}\label{ratio result}
R(\nu)&=&{-\sin(\pi\nu)\Gamma^2(2\nu) \over 
2\sqrt{\pi}\Gamma(2\nu-1)\Gamma(2\nu-1/2)\Gamma(-\nu)}
\nonumber\\
&&\sum_{n=0}^{\infty}{\Gamma(n-\nu)\Gamma(n+1-\nu)\Gamma(\nu+n-1/2)  \over n! \Gamma(n+\nu)\Gamma(n+3/2-\nu)}~,
\end{eqnarray}
which converges as $n^{-\nu-2}$. 
Here, $\nu$ can be treated as a
continuous variable, whereas it has a physical 
meaning when it is a Laughlin fraction $1/m$ ($m$ odd).
At first glance the only physical filling factor 
which one can reach with this series is $\nu=1$. 
Yet, it is possible to  extend $R(\nu)$ to the 
range $[0,1/2]$: the zero frequency noise 
correlations do not contain any true divergence
(it would require the introduction of an
infrared cutoff with a physical origin), 
but the integration 
method which is used here breaks down at $\nu=1/2$, 
a feature which can already be seen in computing 
the average product 
$\langle I_1\rangle\langle I_2\rangle$ 
in a similar manner (although $\langle I_1\rangle$
converges for all $\nu$). It is still possible 
to extract a meaningful result for $\tilde{S}_{12}$
from this integration procedure:
$R(\nu)$ having no poles in $[1/2,1]$, it can 
be analytically continued to the interval $[0,1/2]$. 
Indeed, note that the terms of the series of 
Eq. (\ref{ratio result}) 
are still well defined for $\nu<1/2$). 

The continuation procedure could be
jeopardized if other tunneling operators generated
by the renormalization group (RG) procedure
happened to be more relevant at $\nu=1/2$. 
Consider a higher order tunneling operator  $V_{\vec{n}}
e^{i\sqrt{\nu}\vec{n}.\vec{\phi}}$, 
where $\vec{n}=(n_1,n_2,n_3)$ ($n_l$ integer) satisfies
quasi-particle conservation $\sum_{l=1}^3 n_l=0$ and
$\vec{\phi}=(\phi_1,\phi_2,\phi_3)$  contains the fields of the
three edges. The RG flow is then :  
\begin{equation}
{dV_{\vec{n}}\over dl}
=\left(
1-{\nu\over 2}\sum_{l=1}^3n_l^2 \right) V_{\vec{n}}~.
\label{flow}
\end{equation}
The bare tunneling terms are relevant at
$\nu<1$, and always dominates all other $V_{\vec{n}}$,
which become relevant below $\nu=1/3$ at most. 

For $\nu\simeq 1$, a check
is obtained by direct numerical integration of $S_{12}(t)$.
The comparison between the series solution of  Eq. (\ref{ratio
result}) and the numerical data shows a fair agreement for
$0.7\leq \nu \leq 1$. 

\begin{figure}[htb] \begin{center}
\mbox{\epsfig{file=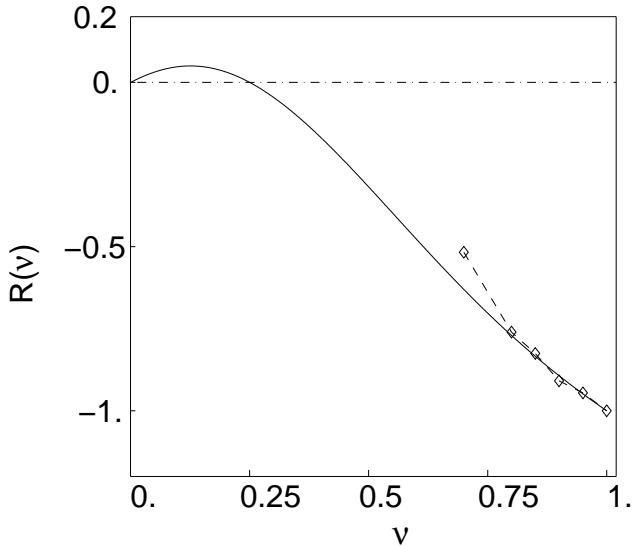,width=8.5cm}} \narrowtext
\caption[to]{Normalized $\omega=0$ 
correlations $R(\nu)$,  
plotted with the analytic expression of Eq. (\ref{ratio result})
for arbitrary $\nu$, and compared to the direct numerical
calculation for $0.7\leq \nu\leq 1$ (lozenges).}  \label{fig3}
\end{center} \end{figure}

Starting from the IQHE and decreasing $\nu$ 
(Fig. \ref{fig3}), the noise correlations between the two
collector edges are reduced in amplitude at any $\nu=1/m$.
When a quasi-particle is detected, in $1$, one is less likely to
observe a depletion of quasi-particles in $2$ than in the case
of noninteracting fermions. The reduction of the (normalized)
noise correlations constitutes  a direct prediction of the
statistical  features associated with fractional quasiparticles
in transport experiments, and should be detectable 
at $\nu=1/3$. 

At $\nu=1/4$, $\tilde{S}_{12}$ vanishes and becomes positive for
lower physical filling factors ($1/5,1/7,...$), which is
reminiscent of bosons  bunching up together.
Positive correlations have
been predicted in superconductor--normal metal junctions
\cite{torres_martin}, and 
bosonic behavior was attributed to the presence of 
Cooper pairs -- effective bosons -- leaking 
on the normal side. 
Here the positive correlations can be either attributed 
to the fact that the fractional statistics are 
bosonic at $\nu\rightarrow 0$, or to the eventual presence 
of composite bosons resulting from attachment of an
odd number of flux tubes \cite{zang_girvin}.
 On the one hand, one is dealing with a
fermionic system where {\it large} negative correlations are
the norm. On the other hand, the presence of
an external magnetic field and the (resulting) collective modes
of the edge excitations favor bosonic behavior. The competition
between these two tendencies yields a statistical signature which
is  close to zero -- analogous to the noise correlations 
of ``classical'' particles. Independent of this sign issue, 
for $\nu\leq 1/3$, the amplitude of the normalized correlations
is strongly reduced and this effect could be checked
experimentally for Laughlin fractions.
The tendency for the noise correlation ratio to
be reduced compared to its non interacting value 
is consistent with the existing data for two-terminal devices
\cite{saminad}, as a connection  exists between the two types of
measurements \cite{martin_landauer,torres_martin}. 
There, shot noise suppression was observed to
be weaker than that of bare electrons, which then
multiplies the shot noise by $1-T$ 
\cite{martin_landauer,reznikov_prl}, the
reflection amplitude. However, a qualitative analysis 
of noise reduction in this situation is rendered difficult 
because of the nonlinear current--voltage characteristics. 
In contrast, an HBT experiment constitutes
a direct and crucial test of the Luttinger liquid models
used to describe the edge excitations in the FQHE,
as it addresses the role of fractional statistics in
transport experiments. These experiments 
could also probe hierachical fractions of the FQHE,
as well as non-chiral Luttinger systems
such as carbon nanotubes.  

We thank H. Saleur for pointing out the relevance of 
the Klein factors. Discussions with D.C. Glattli and
G. Lesovik
are gratefully acknowledged. Two of us (I.S and T.M)
are deeply indebted to their mentors R. Landauer and  
H. J. Schulz.

\end{multicols}
\end{document}